\documentclass[12pt]{article}

\input{epsf}
\def\trace{\mbox{tr}\;}
\def\hsC{hs_{\bf C}(2,2)}
\def\bfdel{\hat{\delta}}

\begin{document}

\begin{titlepage}
\begin{flushright}
NSF-ITP-01-181\\
ITEP-TH-66/01\\
\end{flushright}

\begin{center}
{\Large $ $ \\ $ $ \\
Notes on Higher Spin Symmetries}\\
\bigskip\bigskip\bigskip
{\large Andrei Mikhailov}
\footnote{On leave from the Institute of Theoretical and
Experimental Physics, 117259, Bol. Cheremushkinskaya, 25,
Moscow, Russia.}\\
\bigskip
Institute for Theoretical Physics,\\
University of California, Santa Barbara, CA 93106\\
\vskip 1cm
E-mail: andrei@itp.ucsb.edu
\end{center}
\vskip 1cm
\begin{abstract}
The strong form of the AdS/CFT correspondence implies that the
leading $N$ expressions for the connected correlation functions
of the gauge invariant operators in the free ${\cal N}=4$ supersymmetric
Yang-Mills theory with the gauge group $SU(N)$ correspond
to the boundary S matrix  of the classical interacting theory in
the Anti de Sitter space. It was conjectured recently that the
theory in the bulk should be a local theory of infinitely many
higher spin fields. In this paper we study the
free higher spin fields ($N=\infty$) corresponding to the free scalar
fields on the boundary.
We explicitly construct the
boundary to bulk propagator for the higher spin fields and show
that the classical solutions in the bulk are in one to one
correspondence with the deformations of the free action on the boundary
by the bilinear operators. We also discuss the constraints on the
correlation functions following from the higher spin symmetry.
We show that
the higher spin symmetries fix the correlation functions up to the
finite number of parameters. We formulate sufficient conditions
for the bulk theory to reproduce the free field correlation
functions on the boundary.

\end{abstract}
\end{titlepage}

\section{Introduction.}
\subsection{The problem of interacting higher spin fields.}
In the past few years we have learned a lot about the physics
of the strongly coupled supersymmetric gauge theories from
the study of their supergravity duals \cite{Maldacena,GKP,Witten}.
This progress is based on the AdS/CFT correspondence.
In its basic version the AdS/CFT correspondence
is the equivalence  between the $N=4$
supersymmetric Yang-Mills theory in four dimensions with the gauge
group $SU(N)$ and the Type IIB superstring theory on the space
$AdS_5\times S^5$. The radius of curvature of $AdS_5$ is
equal to the radius of curvature of $S^5$ and is related to the
parameters of the Yang-Mills theory:
\begin{equation}
g_{str}=g_{YM}^2,\;\;\;
\left({R\over l_{str}}\right)^4=g_{YM}^2N
\end{equation}
This equivalence is usually understood as a strong coupling duality,
because the supergravity description of the Type IIB string theory is
valid when $R>>l_{str}$, or $g_{YM}^2N>>1$.
The loop counting parameter on the supergravity side is inversely
proportional to the number of colors $N$.
Therefore the leading large $N$ computations in the gauge theory
correspond to the computations in the classical supergravity
in the bulk.
In the language of string theory this means that we are considering
only the worldsheets with the topology of the sphere.

Suppose that we start decreasing $g_{YM}^2$ keeping $N$ large.
What happens on the string theory side? We should still consider
only the spherical worldsheets; higher genus surfaces will
correspond to $1\over N$ corrections. However the worldsheet
sigma model becomes strongly interacting.
This suggests that from the spacetime point of view
we will be dealing with some complicated classical
theory.

We can study this classical theory by looking at its
boundary S-matrix, which is read from the correlation
functions in the free theory on the boundary. This program was
initiated in  \cite{SHM}.
In the limit $g_{YM}^2=0$ the theory on the boundary becomes the
free ${\cal N}=4$ supersymmetric Yang-Mills.
Consider the correlation functions of $2n$ primary operators
in the free field theory. When $N=\infty$ the correlation
function factorizes into the product of $n$ two point
correlation functions. This means that the theory
in the bulk in this limit should be a free theory.
The fields of this free theory should correspond to the primary
operators in the free theory on the boundary. Since the primary
operators on the boundary are in general tensors of the arbitrary rank,
in the bulk there should be many tensor fields  of the arbitrary
rank.  Perhaps one should think of these higher spin fields as
corresponding to the excited states of the string whose mass
in the limit $R>>l_{str}$ is determined by the radius of
the AdS space rather than the string length.

When $N$ is finite, the correlation functions of the gauge invariant
operators on the boundary do not factorize into the product
of the two point functions. This corresponds to
introducing the interactions in the bulk. It is crucial that
the strength of the interaction depends on $N$ and can be
made arbitrary small by making $N$ very large.
For finite large $N$, the
leading $1\over N$ contribution to the connected correlation
function is given by the tree diagrams. This means that the
correlation functions on the boundary define a classical higher
spin theory
in the bulk, with the coupling constants the positive powers of
$1\over N$.
One can ask whether this theory has local interactions, that is
whether the vertices have a finite number of field derivatives.
If this is true, then it should be possible to determine the
interaction vertices from the symmetries. The resulting local
classical theory would be the effective description of the
string theory in the AdS space of the very small radius.

We do not know any clear string theoretic argument  showing
that the interactions should be local. One can accept this statement
as a hypothesis.

It is remarkable that the problem of constructing the local interactions
of the higher spin fields with the right symmetries is very old.
The question was formulated by C.~Fronsdal \cite{Fronsdal} in 1978.
Significant progress was achieved in the works of E.S.~Fradkin and
M.A.~Vasiliev \cite{FradkinVasiliev} and in the following papers
(see \cite{Vasiliev,SezginSundell} and
references therein.) However the problem of higher
spin interactions remains unsolved.

\subsection{The plan of this paper.}
In our paper we will mostly concentrate on the free higher spin theory
corresponding to $N=\infty$ on the boundary.

{\em Massless fields and gauge symmetries.}
The free field theory has a large group of symmetries involving
the higher derivatives of the fields \cite{hs}.
It is natural to try to find the theory in the bulk which has
this large symmetry group as the group of gauge symmetries.
This group contains scalar gauge symmetries acting on the fields
without derivatives and conformal symmetries acting with one
derivative. These two types of symmetries generate two closed
subgroups, and these are presumably the only symmetries preserved by the
interactions.
In the large radius AdS/CFT correspondence,
the corresponding gauge fields
in the bulk are the vector gauge fields and the spin two gauge field
--- the graviton.
The other symmetries exist only in the free theory. The corresponding
gauge fields in the bulk should be the higher spin gauge fields.

In fact, the correlation functions in the field theory are almost
fixed by the symmetry (we will discuss "almost" in Section 6).
This suggests that if there is  a theory
in the bulk having this large group of symmetries then this theory
should be unique. It would be enough for the purpose of
reproducing the correct correlation functions on the boundary
to have this higher spin symmetry as a global symmetry group.
But we know that it should be a gauge symmetry.
The current operators generating the higher spin symmetry
on the boundary should correspond to the massless fields in the bulk.
For the theory of massless fields to be consistent one needs
gauge invariance. The string theory origin of the higher spin gauge
symmetries is discussed in \cite{Polyakov}.
We will discuss the relation between the global higher
spin symmetry and the gauge transformations for the massless fields in
Section 4.

{\em Consistent truncation.}
The free supersymmetric Yang-Mills theory with the gauge
group $SU(N)$ has $N^2-1$ free gauge fields,
$6(N^2-1)$ free scalars, and $4(N^2-1)$ free complex fermions.
The gauge invariance
implies that we have to consider only the operators consisting
of the traces of the products of the elementary fields.
Consider the operators
which are the products of the traces of the bilinears.
The set of such operators is closed under the operator
product expansion. This suggests
that there is a consistent truncation of the theory in the bulk
to the fields which correspond to the traces of the bilinears.
This statement is a conjecture. For motivation we
use the general interpretation of the OPE in the AdS/CFT
correspondence \cite{OPE}.
Consider $k$ operators ${\cal O}_1\ldots, {\cal O}_k$
inserted at the points $x_1,\ldots,x_k$ on the boundary.
Consider the diagram with $k$ external lines being boundary to bulk
propagators ending
on $x_1,\ldots, x_k$ and one external line being the bulk to bulk
propagator for the bulk field $\phi_{\cal O}$ to the point $z$ inside
AdS:
\begin{center}
\leavevmode
\epsffile{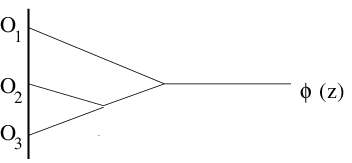}
\end{center}
Here $\phi_{\cal O}$ is the field in the bulk corresponding to
the primary operator ${\cal O}$ on the boundary.
This diagram defines a function $\phi_{\cal O}(z)$ in AdS.
The projection
of $\phi_{\cal O}(z)$ on the space of zero modes of the kinetic operator
for the field $\phi_{\cal O}$ gives a solution $\phi_{\cal O}^{(0)}(z)$
of the free equation
of motion for $\phi_{\cal O}$. This solution corresponds to some operator
on the boundary. This operator has the following meaning: it is
the contribution to the operator product expansion
${\cal O}_1(x_1)\cdots {\cal O}_k(x_k)$ of the single trace
primary operator $\cal O$ and its descendants.
For the free theory on the boundary, we know that the only single
trace primaries in the OPE of the traces of the bilinears are
the traces of the bilinears. This implies that if ${\cal O}$ is not
a bilinear in the free fields, then
\begin{equation}\label{Weaker}
\phi_{\cal O}^{(0)}(z)=0
\end{equation}
The consistent truncation requires that for ${\cal O}(z)$ not a bilinear
in the free fields $\phi_{\cal O}(z)=0$. This is a stronger condition than
(\ref{Weaker}). One could imagine that $\phi_{\cal O}^{(0)}(z)=0$ and
$\phi_{\cal O}(z)\neq 0$. However this is unlikely if the interactions are
local. If $\phi_{\cal O}^{(0)}(z)=0$ and $\phi_{\cal O}(z)\neq 0$, this
most probably implies that the interaction vertex contains the
kinetic operator acting on $\phi_{\cal O}$. But in this case the interaction
vertex can be eliminated by the redefinition of the other interaction
vertices.

The existence of the truncation allows us to consider a simplified problem.
Before looking for the theory in the bulk reproducing the correlation
functions of arbitrary primary operators, one can ask about
the theory describing only the traces of the bilinears. In fact, one can
truncate the theory even further. Consider the bilinears in the free fields
which involve only scalars: $\trace \Phi^I \Phi^J$. There are also bilinears
involving the gauge field and the fermions. But we think that there should
be a consistent truncation to the bilinears involving only the scalars.
The reason is again that all the operators in the operator product expansion
of the product of the bilinears of scalars involve only scalars.

In this paper we will concentrate on this
simplified theory, which describes
only bilinears in scalars.
The currents generating the higher spin symmetries are the traces
of the expressions bilinear in the free fields. In fact, all the primary
fields in the free scalar theory which are bilinears in the free fields
are the conserved currents. Therefore all the operators bilinear in
fields are either conserved currents or their descendants.
The correlation functions of the bilinears have the following
dependence on $N$:
\begin{equation}
\langle
\prod_{k=1}^n \left[{1\over N}\trace \phi(x_k)\phi(y_k)\right]
\rangle
\simeq
{1\over N^{n-2}}
\end{equation}
This means that in the dual theory, vertices with $k$ lines
should count with the coefficient $1\over N^{k-2}$:
\begin{center}
\leavevmode
\epsffile{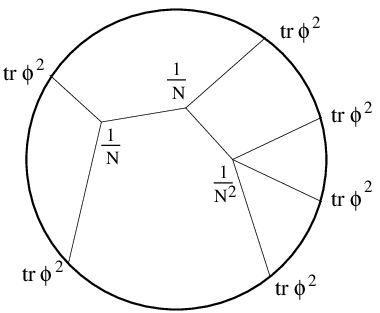}
\end{center}
 Therefore $1\over N$ is the
coupling constant of the theory in the bulk.

{\em Plan.}
In this paper we will discuss the free ($N=\infty$) theory in the bulk.
In Section 2 we will review the structure of the higher spin currents
in the theory of the free scalar fields.
In Section 3 we will study the solutions of the free higher spin
equations in AdS space. We explicitly construct the boundary
to bulk propagator and show that the solutions of the free field
equations are in one to one correspondence with those
deformations of the free field action which are
bilinear in the free fields. This was already done in
\cite{FronsdalDS} but we use a different method.
In Section 4 we discuss the relation
between the equations of motion in the free field theory and
the gauge invariance of the higher spin theory in the bulk.
In Sections 5 and 6 we review the algebraic structure of the higher
spin symmetries and prove that the higher spin symmetries determine
the correlation functions up to the finite number of parameters.
This implies that the higher spin theory
with the right gauge symmetry will automatically reproduce the
correlation functions of the free theory on the boundary.

The group theoretic approach to the theory of free higher spin
fields based on the higher spin symmetries was developed in
the series of papers by E.~Sezgin and P.~Sundell \cite{SezginSundell}.

\section{Primary operators.}
Consider the free complex scalar field
in $D$ dimensions\footnote{In
this paper we
use $d$ for the dimension of the AdS space and $D$ for the dimension
of the boundary.}:
\begin{equation}\label{FreeFieldAction}
S=\int d^Dx\; \partial_i\phi^*\partial^i\phi
\end{equation}
Let us fix a point on the boundary, say $x^i=0$. The action
(\ref{FreeFieldAction}) is invariant under the conformal
transformations. The conformal transformations are the Poincare group
plus the dilatation and the special conformal transformations.
The action of the special conformal transformations on the free
field is:
\begin{equation}
\delta_v\phi = (v\cdot x)(x\cdot\partial)\phi -
{1\over 2}||x||^2 (v\cdot\partial)\phi + {D-2\over 2}(v\cdot x) \phi
\end{equation}
The primaries composed of two scalars are in one to one correspondence
with the traceless symmetric tensors $V^{i_1\ldots i_s}$,
and have the following form:
\begin{equation}\label{Primaries}
\begin{array}{l}
{\cal O}[V]=V^{i_1\ldots  i_s}\sum\limits_{k=0}^s
{(-1)^k\over k!\left(k+{D-4\over 2}\right)!
 (s-k)!\left(s-k+{D-4\over 2}\right)!}
\partial_{i_1}\cdots\partial_{i_k}\phi^*\;
\partial_{i_{k+1}}\cdots\partial_{i_s}\phi,\\[5pt]
g_{i_1 i_2}V^{i_1\ldots i_s}=0
\end{array}
\end{equation}
These primary operators are closely related to the conserved
currents. Let us denote
\begin{equation}
j_{i_1\ldots i_s}=
\sum\limits_{k=0}^s
{(-1)^k\over k!\left(k+{D-4\over 2}\right)!
 (s-k)!\left(s-k+{D-4\over 2}\right)!}
\partial_{i_1}\cdots\partial_{i_k}\phi^*\;
\partial_{i_{k+1}}\cdots\partial_{i_s}\phi-\mbox{traces}
\end{equation}
The operator $j_{i_1\ldots i_s}$ is a primary tensor with the
conformal dimension equal spin plus $D-2$.
The special conformal transformation of
$j^{i_1\ldots i_s}$ is:
\begin{equation}
\delta_v j^{i_1\ldots i_s}=
[D-2+s](v\cdot x) j^{i_1\ldots i_s}+
\sum\limits_{p=1}^s
(v^{i_p}x_k-x^{i_p}v_k)j^{i_1\ldots i_{p-1}k\; i_{p+1}\ldots i_s}
\end{equation}
Then, one can see that
\begin{equation}\label{Conservation}
\partial_{i_1}j^{i_1\ldots i_s}=0
\end{equation}
This is a consequence of the equations of motion
$\partial_i\partial^i\phi=0$. One can prove (\ref{Conservation})
without making calculations.
For any tensor primary $j^{j_1\ldots j_s}$ of
spin $s$ and conformal dimension $D-2+s$ the divergence
$\partial_i j^{ii_2\ldots i_s}$ is again a tensor primary, of
dimension $D-1+s$ \cite{FGG}.
But in the free theory, all the primaries bilinear
 in scalars are given by (\ref{Primaries}) and they
all have conformal dimension equal spin plus $D-2$. Therefore
the divergence of the tensor primary with the
conformal dimension spin plus $D-2$ in the free theory is zero.

The conserved tensor currents are related to the higher
derivative symmetries.
A tensor field
$\xi^{i_2\ldots i_s}$
satisfying the equation
\begin{equation}
\partial^{i_1}\xi^{i_2\ldots i_s}=g^{i_1i_2}\chi^{i_3\ldots i_s}
\;\;\;\;\mbox{(symmetrize $i_1\ldots i_s$)}
\end{equation}
is called a conformal Killing tensor. An example of the conformal
Killing tensor is the product of the conformal Killing vectors,
which are the generators of the conformal symmetries.
Given a conformal Killing tensor, the operator
\begin{equation}
j^{i_1}[\xi]=\xi_{i_2\ldots i_s}j^{i_1\ldots i_s}
\end{equation}
is a conserved current, and generates the symmetry.
We see that the free field theory has infinite dimensional algebra
of symmetries. We will describe the structure of this algebra
for $D=4$ in Section 5.

\section{Higher spin fields.}
\subsection{Free equations for the higher spin fields.}
{\em Free equations and gauge symmetry.}
To the primary operators on the boundary correspond the
higher spin fields in the bulk.
The Lagrangean description of the free higher spin fields was found
by C.~Fronsdal \cite{Fronsdal}.
Spin $s$ fields in the bulk
are rank $s$ tensors, symmetric and double traceless:
\begin{equation}
h^{\mu_1\ldots\mu_s},\;\;\;\mbox{symmetric in}\;\;
\mu_1,\ldots,\mu_s,\;\;\;\\
g_{\mu_1\mu_2}g_{\mu_3\mu_4}h^{\mu_1\ldots\mu_s}=0
\end{equation}
with the gauge transformation:
\begin{equation}
\delta_{\Lambda}h^{\mu_1\ldots\mu_s}=
\nabla^{\mu_1}\Lambda^{\mu_2\ldots\mu_s},
\;\;\;
g_{\mu_2\mu_3}\Lambda^{\mu_2\ldots\mu_s}=0
\end{equation}
The free equations for the higher spin fields are:
\begin{equation}\label{Free}
\begin{array}{l}
\nabla_{\rho}\nabla^{\rho}h_{\mu_1\ldots\mu_s}
-s\nabla_{\rho}\nabla_{\mu_1}h^{\rho}_{\mu_2\ldots\mu_s}+
{1\over 2}
s(s-1)\nabla_{\mu_1}\nabla_{\mu_2}h^{\rho}_{\rho\mu_3\ldots\mu_s}
+\\+
2(s-1)(s+d-3)h_{\mu_1\ldots\mu_s}=0
\end{array}
\end{equation}
The form of the left hand side is fixed by the gauge invariance.
The gauge invariant action for the higher spin fields is known
\cite{Action} but we will not need it here.
It is useful to introduce the de Donder gauge condition:
\begin{equation}\label{deDonderGauge}
F^{\mu_2\ldots\mu_s}[h]=
\nabla^{\rho}h_{\rho\mu_2\ldots\mu_s}-
{s-1\over 2}\nabla_{\mu_2}h^{\rho}_{\rho\mu_3\ldots\mu_s}=0
\end{equation}
The variation of this
gauge condition under the gauge transformation is:
\begin{equation}
\delta_{\Lambda}F^{\mu_2\ldots\mu_s}[h]=\left[
\nabla_{\rho}\nabla^{\rho}+
(s-1)(s+d-3)
\right]
\Lambda^{\mu_2\ldots\mu_s}
\end{equation}
The free
equations (\ref{Free}) are simplified in the de Donder gauge:
\begin{equation}\label{FreeEquationsDD}
\nabla_{\rho}\nabla^{\rho}h^{\mu_1\ldots\mu_s}+
[s^2+(d-6)s-2(d-3)]h^{\mu_1\ldots\mu_s}
+s(s-1)g^{\mu_1\mu_2}h_{\rho}^{\rho\mu_3\ldots\mu_s}=0
\end{equation}
{\em Special gauge for on shell fields.}
Now we want to prove that given a solution $h_{\mu_1\ldots\mu_s}$
of the higher spin equations, one can choose
such a gauge that
\begin{equation}\label{SpecialGauge}
\begin{array}{rl}
1)&
\nabla^{\rho}h_{\rho\mu_2\ldots\mu_s}=0 \\[5pt]
2)& g^{\rho\sigma}h_{\rho\sigma\mu_3\ldots\mu_s}=0
\end{array}
\end{equation}
We start with fixing the de Donder gauge (\ref{deDonderGauge}).
Consider the residual gauge transformations which preserve the
de Donder gauge:
\begin{equation}\label{EqnForLambda}
\begin{array}{c}
\nabla_{\rho}\nabla^{\rho}\Lambda^{\mu_2\ldots\mu_s}+
m^2\Lambda^{\mu_2\ldots\mu_s}=0,\\[5pt]
\mbox{with}\;\;m^2=(s-1)(s+d-3)
\end{array}
\end{equation}
Then $\nabla_{\rho}\Lambda^{\rho\mu_3\ldots\mu_s}$ satisfies:
\begin{equation}\label{EqnDivLambda}
\begin{array}{c}
\nabla_{\rho}\nabla^{\rho}
(\nabla_{\sigma}\Lambda^{\sigma\mu_3\ldots\mu_s})+
\tilde{m}^2\nabla_{\sigma}\Lambda^{\sigma\mu_3\ldots\mu_s}=0
\\[5pt]
\mbox{with}\;\;\tilde{m}^2=s^2+(d-2)s-2
\end{array}
\end{equation}
One can see from (\ref{FreeEquationsDD}) that
$h_{\sigma}^{\sigma\mu_3\ldots\mu_s}$ satisfies the same
equation\footnote{One can see it without making any calculations.
The equation for $\Lambda$ preserving the
de Donder gauge condition is a second order differential equation.
From this equation follows the equation for the divergence of
$\Lambda$ which is again a second order differential equation.
But the divergence of $\Lambda$ will automatically satisfy the
equation of motion for the trace of $h$.} (\ref{EqnDivLambda}):
\begin{equation}\label{EqnForTrace}
\nabla_{\rho}\nabla^{\rho}
h_{\sigma}^{\sigma\mu_3\ldots\mu_s}+
\tilde{m}^2 h_{\sigma}^{\sigma\mu_3\ldots\mu_s}=0
\end{equation}
We want to prove that any solution
$f^{\mu_3\ldots\mu_s}=h_{\rho}^{\rho\mu_3\ldots\mu_s}$
of (\ref{EqnForTrace}) is
obtained from a solution of (\ref{EqnForLambda}) as
\begin{equation}
f^{\mu_3\ldots\mu_s}=\nabla_{\sigma}\Lambda^{\sigma\mu_3\ldots\mu_s}
\end{equation}
This $\Lambda$ will be then the residual gauge transformation removing the
trace of $h$.
We will need first the following statement: for any $f^{\mu_3\ldots\mu_s}$
(arbitrary $f^{\mu_3\ldots\mu_s}$,
does not have
to satisfy any equation)
we can find such a traceless $\Lambda^{\mu_2\ldots\mu_s}$
that:
\begin{equation}\label{FViaLambda}
f^{\mu_3\ldots\mu_s}=\nabla_{\sigma}\Lambda^{\sigma\mu_3\ldots\mu_s}
\end{equation}
This is proven in the Appendix.
Now suppose that $f$ satisfies the equation (\ref{EqnForTrace}) and
find some $\Lambda$ so that (\ref{FViaLambda}). We want to prove
that in fact $\Lambda$ satisfies (\ref{EqnForLambda}).
Let us decompose
\begin{equation}
\Lambda=\Lambda_0+(\nabla_{\rho}\nabla^{\rho}+m^2)\Xi
\end{equation}
where $\Lambda_0$ is in the kernel of
$\nabla_{\rho}\nabla^{\rho}+m^2$. Then
\begin{equation}
f=\nabla\cdot\Lambda=\nabla\cdot\Lambda_0+
(\nabla_{\rho}\nabla^{\rho}+\tilde{m}^2)\nabla\cdot\Xi
\end{equation}
But $f$ is in the kernel of
$\nabla_{\rho}\nabla^{\rho}+\tilde{m}^2$, therefore
$\nabla\cdot\Xi$ is in the kernel of
$\nabla_{\rho}\nabla^{\rho}+\tilde{m}^2$. This implies that
\begin{equation}
f=\nabla\cdot\Lambda=\nabla\cdot\Lambda_0
\end{equation}
This is what we wanted to prove.

{\em The action.} There is a gauge invariant action for
the higher spin fields in AdS space, but we don't need it here.
The equation (\ref{FreeEquationsDD}) for traceless $h$ can be obtained
from the following action:
\begin{equation}\label{Action}
S=(-1)^s\int \left(
-\nabla^{\mu}h^{\mu_1\ldots\mu_s}\nabla_{\mu}h_{\mu_1\ldots\mu_s}
+
[s^2+(d-6)s-2(d-3)]h^{\mu_1\ldots\mu_s}h_{\mu_1\ldots\mu_s}\right)
\end{equation}
Our convention for the metric is that the signature is mostly minus.
In Euclidean AdS, the metric is negative definite.
The positivity of the action at least for $s\geq 3$ and 
$d\geq 3$ 
can be 
proven by the following trick. Define
$\tilde{h}^{\mu_1\ldots\mu_s}$:
$$
h^{\mu_1\ldots\mu_s}=
\phi \tilde{h}^{\mu_1\ldots\mu_s}
$$
Here $\phi$ is a solution to the Laplace equation
$$
\partial_{\mu}\partial^{\mu}\phi={(d-1)^2\over 4}\phi
$$
In Poincare coordinates, $\phi=z^{d-1\over 2}$.
After integration by parts, the action becomes
\begin{equation}
\begin{array}{c}
S=(-1)^s\int \phi^2\left(
-\nabla^{\mu}\tilde{h}^{\mu_1\ldots\mu_s}\nabla_{\mu}
\tilde{h}_{\mu_1\ldots\mu_s}
+\right.\\[5pt]\left.+
\left[{(d-1)^2\over 4}+s^2+(d-6)s-2(d-3)\right]
\tilde{h}^{\mu_1\ldots\mu_s}\tilde{h}_{\mu_1\ldots\mu_s}\right)
\end{array}
\end{equation}
which is manifestly positive for $s\geq 3$, $d\geq 3$.
\subsection{From the boundary to the bulk.}
The metric on AdS space has a double pole at the boundary.
This means that one can find ``defining'' function $r$ which
has a simple zero at the boundary, so that
$\overline{g_{..}}=r^2g_{..}$
is nonsingular. The boundary of AdS space can be defined as a set of
equivalence classes of sequences $\{x_i|i=1,\ldots,\infty\}$ which
are Cauchy sequences with respect to the metric $\overline{g_{..}}$
and are not convergent to anything inside $AdS$. A vector field in
AdS space acts on Cauchy sequences, and therefore on the boundary.
This defines a restriction of the vector fields in the bulk to the
vector fields on the boundary.

We will find that for any symmetric traceless tensor field
$V^{i_1\ldots i_s}(x)$ on the boundary there is a solution
$h[V]$ of the free higher spin equations in the bulk such that
${1\over r^{2}}h[V]$ has a restriction on the boundary equal to
$V^{i_1\ldots i_s}$. Schematically,
\begin{equation}\label{Restriction}
\lim\limits_{v\to x}{1\over r^2}h[V]^{i_1\ldots i_s}=
V(x)^{i_1\ldots i_s}
\end{equation}
We will also prove that if the restriction of the solution
${1\over r^2}h^{\mu_1\ldots\mu_s}$
to the boundary is zero than $h^{\mu_1\ldots\mu_s}$ is zero.

These results are expected from the fact that the space of the
solutions of the free equations for the higher
spin fields in the bulk of the AdS space is the tensor product
of the two doubleton representations \cite{FronsdalDS}.

Let us think of the AdS space as the hyperboloid embedded into
${\bf R}^{2+(d-1)}$. This is the set of vectors $v$ in
${\bf R}^{2+(d-1)}$
with the length square $||v||^2=1$. The quadratic
form has the signature 2 pluses and $d-1$ minuses.
The boundary of this hyperboloid is the projectivization of
the light cone;
the point of the boundary is a lightlike vector $l$, modulo the rescaling.
We will think of symmetric traceless tensors on the boundary as symmetric
traceless tensors in the tangent space to the light cone, modulo the generator
of the cone:
\begin{equation}\label{VDefinition}
\begin{array}{ll}
V^{I_1\ldots I_s}=V^{(I_1\ldots I_s)},\;\;\;&
\mbox{($V$ is symmetric tensor in ${\bf
R}^{2+(d-1)}$)}
\\[5pt]
l_{I}V^{II_2\ldots I_s}=0, \;\;\; V_I^{I I_3\ldots I_s}=0\;\;\;&
\mbox{($V$ is tangent to the light cone at point $l$ and
traceless)}
\\[5pt]
V^{I_1\ldots I_s}\equiv V^{I_1\ldots I_s}+
l^{I_1}W^{I_2\ldots I_s} \;\;\;&
\mbox{($V$ is defined modulo the generator of
the cone)}
\end{array}
\end{equation}
Here the capital latin indices label the
tangent space to ${\bf R}^{2+(d-1)}$.
The space of such $V$ with this equivalence
relation is the same as the space of traceless symmetric
tensors in the tangent
space to the boundary at the point $l$.  For each point $v$
on the hyperboloid,
we consider a map from the tangent space to ${\bf R}^{2+(d-1)}$ to
the tangent
space of the hyperboloid:
\begin{equation}
\hat{\delta}_I^{\mu}=\delta_I^{\mu}-{l^{\mu} v_I\over (v\cdot l)}
\end{equation}
We use the greek indices to label the tangent space to $AdS_d$.
Consider the following tensor field on $AdS_d$:
\begin{equation}
h_V^{\mu_1\ldots\mu_s}=
{1\over (v\cdot l)^{d-3+s}}
\hat{\delta}_{I_1}^{\mu_1}\cdots\hat{\delta}_{I_s}^{\mu_s}
V^{I_1\ldots I_s}
\end{equation}
This tensor field does not change if we add to
$V^{I_1\ldots I_s}$ the expression proportional to the generator
of the light cone; therefore it is correctly defined as a function
of $V$ modulo the equivalence relations (\ref{VDefinition}).
It transforms
with the weight $-(d-3+s)$ under the rescalings of $l$.
Notice that $h_V$ is
invariant under the special conformal transformations preserving
the point $l$.
One can show that $h_V$ satisfies the free equation (\ref{Free}).

Tensor densities of weight $\Delta$ are the quantities
which transform as tensors under diffeomorphisms and get rescaled
by the factor $\lambda^{\Delta}$ under the rescaling of the metric
$g_{ij}\to \lambda^{2\Delta}g_{ij}$.
We see that $V\mapsto h_V$ is a map from the space of symmetric traceless
tensor densities $V$ of the weight $d-3+s$ at the point $l$ of the boundary
to the space of the solutions of (\ref{Free}) in the bulk.
Let us introduce the notation:
\begin{equation}
G_l(v)^{\mu_1\ldots\mu_s}_{I_1\ldots I_s}=
{1\over (v\cdot l)^{d-3+s}}
\left[
\hat{\delta}_{I_1}^{\mu_1}\cdots\hat{\delta}_{I_s}^{\mu_s}
-\mbox{traces}_{(I)}\right]
\end{equation}
Here $\mbox{traces}_{(I)}$ means that we subtract expressions
proportional to $g^{I_jI_k}$ so that the resulting expression
is symmetric and traceless in $I_1\ldots I_s$.
This is the boundary to bulk propagator for the higher spin fields.
The solution of our problem (\ref{Restriction}) is given by:
\begin{equation}\label{Integrated}
h[V]^{\mu_1\ldots\mu_s}={1\over {\cal N}(s,d)}
\int d^{d-1}l\; G_{l}(v)_{I_1\ldots I_s}^{\mu_1\ldots\mu_s}
V(l)^{I_1\ldots I_s}
\end{equation}
where
\begin{equation}
{\cal N}(s,d)=2^s\; \pi^{d-1\over 2}
\Gamma\left({d-5\over 2}+s\right)
\sum\limits_{p=0}^{s}(-1)^p
{s!(d-4+s)!\over (s-p)! (d-4+s+p)!}
\end{equation}
Let us prove this statement.
Since we study the behaviour of the solution near the boundary,
it is useful to introduce the Poincare coordinates.
In the Poincare coordinates, the metric has the form
\begin{equation}\label{PoincareMetric}
ds^2={1\over (z^0)^2}\left((dz^0)^2+(dz^i)^2\right)
\end{equation}
The vector $v$ on the hyperboloid and the scalar product $(v\cdot l)$
are in Poincare coordinates:
\begin{equation}\label{Poincare}
v=\left[{z_0^2+z_i^2+1\over 2z_0},
        {z_0^2+z_i^2-1\over 2z_0},
        {z_i\over z_0}\right],\;\;\;, l=[1,-1,\vec{0}],\;\;\;
(v\cdot l)={z_0^2+z_i^2\over z_0}
\end{equation}
Let us write the boundary to bulk propagator in the Poincare
coordinates. It is straightforward to compute $\bfdel$ in
Poincare coordinates using the formula:
\begin{equation}
\bfdel^{\mu}_I=(v\cdot l)\partial^{\mu}\left[{1\over (v\cdot l)}
{\partial\over\partial l^I} (v\cdot l)\right]
\end{equation}
The result is:
\begin{equation}
G_l(v)_{\mu_1\ldots\mu_s}^{i_1\ldots i_s}
=2^s\left[{z_0\over z_0^2+\vec{z}^2}\right]^{d-3}
\left[\partial_{\mu_1}{z^{i_1}\over z_0^2+\vec{z}^2}
\cdots\partial_{\mu_s}{z^{i_s}\over z_0^2+\vec{z}^2}
\right]
\end{equation}
The tensor field on the boundary $V(x^i)^{i_1\ldots i_s}$.
depends on $x^i$. But at small $z_0$ the
main contribution to the integral in (\ref{Integrated}) is from the
region close to $z^i=x^i$. Therefore,
one can approximate the integral (\ref{Integrated}) with the
constant $V^{i_1\ldots i_s}$.
We want to integrate the traceless part:
\begin{equation}
\begin{array}{l}
\int d^{d-1}\vec{z} G_{\vec{z}}(v)^{j_1\ldots j_s}_{i_1\ldots i_s}
=\\[5pt]=
2^sz_0^{d-3}\int d^{d-1}\vec{z}\sum\limits_{p=0}^s
{C_s^p\over (d-3+s)\cdots (d-4+p+s)}
\delta_{i_{p+1}}^{j_{p+1}}\cdots\delta_{i_s}^{j_s}
z^{j_1}\cdots z^{j_p}\partial_{i_1}\cdots\partial_{i_p}
{1\over (z_0^2+\vec{z}^2)^{d-3+s}}
=\\[5pt]=
2^sz_0^{d-3}\int d^{d-1}\vec{z}\sum\limits_{p=0}^s
{(-1)^p p!C_s^p\over (d-3+s)\cdots (d-4+p+s)}
\delta_{i_{1}}^{j_{1}}\cdots\delta_{i_s}^{j_s}
{1\over (z_0^2+\vec{z}^2)^{d-3+s}}
=\\[5pt]=
2^s\; z_0^{2-2s}\;\pi^{d-1\over 2}
\Gamma\left({d-5\over 2}+s\right)
\sum\limits_{p=0}^{s}(-1)^p
{s!(d-4+s)!\over (s-p)! (d-4+s+p)!}
\delta_{i_{1}}^{j_{1}}\cdots\delta_{i_s}^{j_s}
\end{array}
\end{equation}
(we have taken the derivatives and neglected the terms proportional
to $g_{i_pi_q}$ and $g_{j_pj_q}$.)
In this formula the upper indices are in the tangent space to the
boundary and the lower indices are in the tangent space to the
AdS (indices parallel to the boundary in the Poincare coordinates).
Rising indices with the inverse metric tensor proportional to $z_0^2$
we get the right asymptotic behaviour (\ref{Restriction}).

We have proven that $G_{l}(v)_{I_1\ldots I_s}^{\mu_1\ldots\mu_s}$
is the boundary to bulk propagator for the higher spin fields.
It is divergenceless and traceless:
\begin{equation}
\begin{array}{l}
\nabla^{\rho}G^{i_1\ldots i_s}_{\rho\mu_2\ldots\mu_s}=0
\\[5pt]
g^{\rho\sigma}h^{i_1\ldots i_s}_{\rho\sigma\mu_3\ldots\mu_s}=0
\end{array}
\end{equation}
These conditions are the special gauge
conditions (\ref{SpecialGauge}) for the on shell fields.

\subsection{From the bulk to the boundary.}
For the symmetric tensor field on the boundary we have
constructed the symmetric traceless field in the bulk
which is the solution to the free higher spin equations
of motion.
Here we will describe the inverse map.

The inverse map is the restriction of the tensor field
in the bulk to the boundary. This restriction is defined
(in the Poincare coordinates) as follows:
\begin{equation}
\lim_{x\to bnd.}r^{-2}h^{I_1\ldots I_s}(x)
\end{equation}
{\em The restriction to the boundary is traceless.}
Let us consider the divergence of $h$ with the indices
parallel to the boundary:
\begin{equation}
\begin{array}{rl}
0&=\nabla_{\rho}h^{\rho I_2\ldots I_s}
=\\[5pt]&=
z^{d+2(s-1)}\partial_{\rho}\left( z^{-d-2(s-1)}h^{\rho I_2\ldots I_s}
\right)=\partial_z h^{zI_2\ldots I_s}
+\partial_I h^{II_2\ldots I_s}
-{d+2(s-1)\over z} h^{zI_2\ldots I_s}
\end{array}
\end{equation}
Therefore
\begin{equation}\label{HzI}
h^{zI_2\ldots I_s}=-z^{d+2s-2}\int_{z_0(x)}^zd\tilde{z}\;
\tilde{z}^{-(d+2s-2)}\partial_I h^{II_2\ldots I_s}(\tilde{z})
\end{equation}
When we go to the boundary $h^{I_1\ldots I_s}$ behaves
like $z^{2}$, then (\ref{HzI}) tells us that $h^{zI_2\ldots I_s}$
goes like $z^{3}$. Now let  us use the equation for
$(\nabla\cdot h)^{z\mu_3\ldots\mu_s}=0$:
\begin{equation}
\begin{array}{rl}
0&=\nabla_{\rho} h^{\rho z I_3\ldots I_s}
 =\\[5pt]&=
\partial_z h^{zzI_3\ldots I_s}+\partial_I h^{zII_2\ldots I_s}-
{1\over z}[d+2s-2]h^{zzI_3\ldots I_s}
\end{array}
\end{equation}
We have used the tracelessness of $h$. Therefore
$h^{zzI_3\ldots I_s}$ goes as $z^{4}$. Since $h^{I_1\ldots I_s}$ goes
as $z^{2}$ and $h^{\mu_1\ldots\mu_s}$ is traceless, this implies
that the restriction of $h$ to the boundary is traceless.

{\em The restriction of the nonzero solution is nonzero.}
Let us prove it. First let us notice that $h$ cannot vanish
at the infinity as fast or faster than $z^{d-2+2s}$.
Indeed, if it vanished that fast we could integrate by parts
in the action and prove that the action is zero. Which contradicts
to the positivity of the action. If $h$ vanishes slower than
$z^{d-2+2s}$ then the $h^{I_1\ldots I_s}$ component of $h$ is
much larger near the boundary than the other components.
The action (\ref{Action}) near the boundary is:
\begin{equation}
\begin{array}{l}
S=\int d^{d-1}\vec{z}\; \int {dz\over z^d}\left[
z^2\partial_z(z^{-s}h^{I_1\ldots I_s})
   \partial_z(z^{-s}h^{I_1\ldots I_s})+\right.
\\[5pt]+\left.
[s^2+(d-5)s-2(d-3)]z^{-2s}h^{I_1\ldots I_s}h^{I_1\ldots I_s}
\right]
\end{array}
\end{equation}
The resulting equations of motion have two solutions:
\begin{equation}
\begin{array}{l}
h^{I_1\ldots I_s}(z)=z^2 V^{I_1\ldots I_s},\;\;
\\[5pt]
h^{I_1\ldots I_s}(z)=z^{2s+d-3} V^{I_1\ldots I_s}
\end{array}
\end{equation}
This means that if the leading component of $h^{I_1\ldots I_s}$
decreases near the boundary faster than $z^2$ than it decreases
as fast as $z^{2s+d-3}$. But if it decreases as fast as $z^{2s+d-3}$
then it is zero (because we can integrate by parts in the action
and prove that the action is zero; but the action is positive definite.)

Therefore the correspondence between the traceless tensor fields on
the boundary and the solutions of the massless higher spin equations
in the bulk is one to one.

\section{More on gauge transformations.}
\subsection{The divergence of the current on the boundary is the
gauge transformation in the bulk.}
The equations of motion in the free theory imply that the primary
currents are conserved. In this section we will show that the
divergence of the boundary to bulk propagator is a pure gauge in the bulk:
\begin{equation}
\partial_i\; G^{ii_2\ldots i_s}_{\mu_1\ldots\mu_s}=
\nabla_{\mu_1}\Lambda_{\mu_2\ldots\mu_s}^{i_2\ldots i_s}
\end{equation}
(latin indices are in the tangent space to the boundary, and greek indices
in the bulk; symmetrization of the greek indices).
We will use the Poincare coordinates (\ref{PoincareMetric}),
(\ref{Poincare}). Let us denote
\begin{equation}\label{DivNotations}
\begin{array}{l}
\Phi:={1\over (v\cdot l)^2}={z_0^2\over (z_0^2+\vec{z}^2)^2}
\\[5pt]
J_i=\Phi^{-1}\partial_i \Phi
\end{array}
\end{equation}
The boundary to bulk propagator is:
\begin{equation}
G_l(v)_{i_1\ldots i_s}^{\mu_1\ldots\mu_s}=
\left(-{1\over 4}\right)^s\Phi\;
\partial^{\mu_1}J_{i_1} \cdots \partial^{\mu_s}J_{i_s}
-\mbox{traces}
\end{equation}
Using $
\partial_i J_j=-4\delta_{ij}\Phi+{1\over 2}J_iJ_j-
{1\over 4}\delta_{ij}J_k^2
$
one can see that
\begin{equation}\label{DivOfPropagator}
\begin{array}{l}
\partial^i\left(
\Phi^{d-3\over 2} \partial^{\mu_1}J_i
\partial^{\mu_2}J_{i_2}\cdots
\partial^{\mu_s}J_{i_s}-\mbox{traces}_{(i)}
\right)
=\\[5pt]=
-4{2s^2+(3d-11)s+(d^2-7d+12)\over 2s+d-5}
\left(\Phi^{d-3\over 2}
\partial^{\mu_1}\Phi\partial^{\mu_2}J_{i_2}\cdots
\partial^{\mu_s}J_{i_s}
 -\mbox{traces}_{(i)}\right)
\end{array}
\end{equation}
It is convenient to rewrite the right hand side
in terms of the Killing vector fields. The Killing vector fields
corresponding to the special conformal transformations leaving the
point $l$ on the boundary can be expressed through $\Phi$:
\begin{equation}
{\bf V}_{i}^{\mu}=(v\cdot l)\hat{\delta}_{i}^{\mu}
={1\over \Phi}\partial^{\mu}J_i
\end{equation}
For $i$ going from $1$ to $4$ these vector fields generate
the special conformal transformations preserving the point $l$
on the boundary.
With the help of these Killing vector fields
we can show that the right hand side of
(\ref{DivOfPropagator}) is a pure gauge:
\begin{equation}
\begin{array}{l}
\Phi^{d-3\over 2}\partial^{\mu_1}\Phi \Phi^{s-1}
 {\bf V}^{\mu_2}_{i_2}\cdots{\bf V}^{\mu_s}_{i_s}
-\mbox{traces}_{(i)}=\\[5pt]=
{2\over 2s+d-3}\nabla^{\mu_1}\left( \Phi^{s+{d-3\over 2}}
{\bf V}^{\mu_2}_{i_2}\cdots{\bf V}^{\mu_s}_{i_s}
-\mbox{traces}_{(i)}\right)
\end{array}
\end{equation}
We see that the divergence of the current on the boundary corresponds
to the total gauge in the bulk:
\begin{equation}\label{DivergenceIsGauge}
\begin{array}{l}
\partial^i G_{ii_2\ldots i_s}^{\mu_1\ldots\mu_s}=
2\left(-{1\over 4}\right)^{s-1}
{s+d-4\over 2s+d-5}
\times\\[5pt]\times
\nabla^{\mu_1}\left(\Phi^{s+{d-3\over 2}}{\bf V}_{i_2}^{\mu_2}
\cdots{\bf V}_{i_s}^{\mu_s}-\mbox{traces}\right)
=\\[5pt]=
2\left(-{1\over 4}\right)^{s-1}
{s+d-4\over 2s+d-5}
\nabla^{\mu_1}\left[{1\over (v\cdot l)^{s+d-2}}
(\hat{\delta}^{\mu_2}_{i_2}\cdots \hat{\delta}^{\mu_s}_{i_s}
-\mbox{traces})\right]
\end{array}
\end{equation}
This formula provides us with the correspondence
between the conformal Killing tensors on the boundary and the
traceless Killing tensors in the bulk.
Let us consider the conformal Killing tensor on the boundary, which is
a tensor field $\xi_{i_2\ldots i_s}$ such that
\begin{equation}
\partial_{i_1}\xi_{i_2\ldots i_s}=g_{i_1 i_2}\eta_{i_3\ldots i_s}
\end{equation}
Integrating by parts and using the definition of the conformal
Killing tensor we find:
\begin{equation}
\int d^{d-1}l \xi^{i_2\ldots i_s}
\partial^i G_{ii_2\ldots i_s}^{\mu_1\mu_2\ldots \mu_s}=0
\end{equation}
Comparing this to (\ref{DivergenceIsGauge}) we find that for any
conformal Killing tensor $\xi$ on the boundary the tensor
field $\Lambda[\xi]$ in the bulk defined as
\begin{equation}\label{KillingsCorrespondence}
\Lambda[\xi]^{\mu_2\ldots\mu_s}:=
\int d^{d-1}l\; \xi(x)^{i_2\ldots i_s}
{1\over (v\cdot l)^{s+d-2}}(\hat{\delta}_{i_2}^{\mu_2}
\cdots \hat{\delta}_{i_s}^{\mu_s}-\mbox{traces}_{(i)})
\end{equation}
is a traceless Killing tensor field. The integral
(\ref{KillingsCorrespondence}) is the correspondence between the conformal
Killing tensors on the boundary and the traceless Killing tensors
in the bulk.
The inverse map is, in Poincare coordinates
 (up to coefficient):
\begin{equation}
\Lambda^{\mu_2\ldots\mu_s}\to
\lim\limits_{z_0\to 0} \Lambda^{i_2\ldots i_s}
\end{equation}
The  higher derivative symmetries on the boundary
correspond to the global gauge symmetries in the bulk.

\subsection{Triple interactions and the deformations of the gauge
transformations.}
Suppose that we have constructed such a triple interaction that
it is invariant on shell under the gauge transformations, and
reproduces the right correlation functions on the boundary. In the theory
we are looking for such a triple interaction cannot possibly
be invariant off shell.
The variation of the triple interaction under the gauge transformation
will be proportional to the action of the kinetic operator
on the external line, so that it vanishes on shell. This is compensated
by the appropriate deformation of the gauge transformation:
\begin{equation}
\delta_{\Lambda}=\delta^{(0)}_{\Lambda}+\lambda\delta^{(1)}_{\Lambda}
+\ldots
\end{equation}
where $\lambda={1\over N}$ is the coupling constant.
We should be able to choose $\delta^{(1)}_{\Lambda}h$ a linear
operator acting on $h$ so that the variation of the kinetic term in
the action under $\delta^{(1)}_{\Lambda}$ compensates for the variation
of the triple interaction vertex under $\delta^{(0)}_{\Lambda}$.
We cannot say much about $\delta^{(1)}_{\Lambda}$ because we have not
solved the problem of interactions. Here we will explain a general
property of $\delta^{(1)}_{\Lambda}$:
for $\Lambda$ the Killing tensor, the action of
$\delta^{(1)}_{\Lambda}$ on the solutions of the free equation of motion
corresponds to the action of the corresponding
higher derivative symmetry on the
operators on the boundary. This, in particular, shows that
$\delta^{(1)}_{\Lambda}$ cannot be zero.

Let us formulate this more precisely.
Notice that for $\Lambda$ being a Killing tensor, the triple interaction
is invariant under $\delta^{(0)}_{\Lambda}$
(just because $\delta^{(0)}_{\Lambda}$ is zero for Killing $\Lambda$.)
Therefore $\delta^{(1)}_{\Lambda}$ for Killing $\Lambda$ should
be a symmetry of the free action. In particular, it should transform
the solutions of the free equations into the solutions.
The solutions of the free equations correspond to the primary operators
on the boundary. Given the primary operator ${\cal O}$ we will
denote $h_{\cal O}$ the corresponding solution of the higher spin
equations.  We want to show that
\begin{equation}\label{ActionOnShell}
\delta^{(1)}_{\Lambda[\xi]}h_{\cal O}=h_{\delta_{\xi}{\cal O}}
\end{equation}
Here $\Lambda[\xi]$ is the traceless Killing tensor field in the bulk
corresponding to the conformal Killing tensor field $\xi$ on
the boundary, and $\delta_{\xi}{\cal O}$ is the transformation
of the boundary operator ${\cal O}$ under the higher derivative
symmetry corresponding to the conformal Killing tensor $\xi$.
Indeed, let us consider the following correlation function:
\begin{equation}
\langle {\cal O}_1(x_1) \oint_{C} *(\xi_{i_2\ldots i_s}
j^{i i_2\ldots i_s}) {\cal O}_2(x_2) \rangle
\end{equation}
where the contour $C$ is around the point $x_2$. Naively this integral
is zero because the divergence of the current is zero and we can
contract the contour. But we know it is not zero because of the
singularity in the OPE of $j$ and ${\cal O}_2$. In fact, it is
equal to
\begin{equation}\label{EqualTwoPoint}
\langle {\cal O}_1(x_1)\delta_{\xi}{\cal O}_2(x_2)\rangle
\end{equation}
Let us see what happens when we compute this correlator on the
AdS side. Let us use the Stokes theorem
$$\oint_C *(\xi_{i_2\ldots i_s}
j^{i i_2\ldots i_s})=\int_D d^4 x \partial_i j^{i i_2\ldots i_s}
\xi_{i_2\ldots i_s}$$
The divergence of the current on the boundary gives pure gauge in the
bulk. If the triple interaction was invariant under the gauge
transformation, then the correlation function would be zero.
But in fact the gauge variation gives the expression proportional
to the equation of motion for $h_{{\cal O}_1}$ plus the term proportional
to the equation of motion for $h_{{\cal O}_2}$. The first term gives zero
contrubution because $\Lambda$ goes to zero near the point of the
boundary where ${\cal O}_1$ is inserted. But the term proportional
to $h_{{\cal O}_2}$ gives nonzero result because of the boundary term.

Let us see how this boundary term is compenstated by the variation
of the free action. Notice that the variation of the free action
under $\delta^{(0)}_{\Lambda}$ is zero, because $\Lambda$ goes either
to zero or to the Killing vector near the insertion points.
Therefore we have to consider only the variation under
$\delta^{(1)}_{\Lambda}$. It consists of two terms:
\begin{equation}\label{VariationOfSFree}
\delta^{(1)}_{\Lambda} S_0[h_{{\cal O}_1}, h_{{\cal O}_2}]
=S_0[\delta^{(1)}_{\Lambda}h_{{\cal O}_1}, h_{{\cal O}_2}]
+S_0[h_{{\cal O}_1}, \delta^{(1)}_{\Lambda}h_{{\cal O}_2}]
\end{equation}
We find it very convenient for this type of calculations
to replace the insertion of the operator on the boundary
by some source in the bulk, localized near the boundary.
Let us consider the boundary to bulk propagator $G_l(v)$ and
modify it by multiplying by the step function:
\begin{equation}
G_l(v)_{reg}=\theta[(v\cdot l)>\epsilon]G_l(v)
\end{equation}
This modification corresponds to replacing the boundary to
bulk propagator with the bulk to bulk propagator contracted
with the "double layer" source localized at $(v\cdot l)=\epsilon$. When $\epsilon$
goes to zero we return to the ordinary boundary to bulk propagator.
\begin{center}
\leavevmode
\epsffile{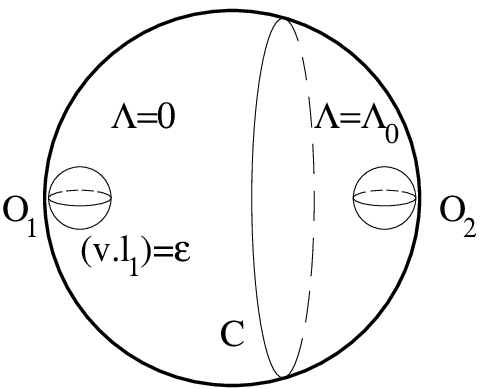}
\end{center}
The second term in (\ref{VariationOfSFree}) is zero.
Indeed, we can integrate by parts
and use the $h_{{\cal O}_1}$ equation of motion; there is no boundary
term because $\Lambda$ goes to zero near the source of
$h_{{\cal O}_1}$. The first term after integration by parts
gives nonzero result from the source of $h_{{\cal O}_2}$.
It depends on $\delta^{(1)}_{\Lambda}h_{{\cal O}_1}$
near the source of $h_{{\cal O}_2}$.
We are assuming that the variation
$\delta_{\Lambda}^{(1)}h_{{\cal O}_1}$ is given
by a local expression in $h_{{\cal O}_1}$ and $\Lambda$.
But near the insertion point of ${\cal O}_2$ $\Lambda$
becomes a Killing tensor. Therefore the boundary term is equal
to the variation of the free action
$S_0[h_{{\cal O}_1}, h_{{\cal O}_2}]$ under the transformation
of $h_{{\cal O}_1}$ by the Killing $\Lambda$.

Comparing (\ref{EqualTwoPoint}) and
(\ref{VariationOfSFree}) we see that the free action plus the cubic
interaction is invariant only if
\begin{equation}
S_0[\delta^{(1)}_{\Lambda}h_{{\cal O}_1}, h_{{\cal O}_2}]+
\langle {\cal O}_1(x_1)\delta_{\xi}{\cal O}_2(x_2)\rangle=0
\end{equation}
for arbitrary ${\cal O}_1$ and ${\cal O}_2$, which implies
(\ref{ActionOnShell}).

Assume that the following is true:
\begin{enumerate}
\item{
The higher spin theory in the bulk gives the
three point functions on the boundary equal to the three point
functions of the free field theory.}
\item{For Killing $\Lambda$, $\delta^{(1)}_{\Lambda}$ is the symmetry
of the action. In other words, $\delta^{(2)}_{\Lambda}=
\delta^{(3)}_{\Lambda}=\ldots=\delta^{(n>1)}_{\Lambda}=0$ for
Killing $\Lambda$.}
\end{enumerate}
Then the higher spin symmetries of the free theory are
the symmetries of the boundary S-matrix in the higher spin theory
in AdS, and the action of these symmetries on the asymptotic
states agrees with the action on the primary operators on the
boundary, equation (\ref{ActionOnShell}).
We will show in the next section that at least for the four
dimensional boundary the higher spin symmetries
fix the correlation functions up to a finite number of parameters
 (corresponding to the permutations of the doubletons).
These parameters will presumably be fixed by the positions of
the singularities in the correlation functions. This means that
if the singularities are correctly reproduced, then the theory
in the bulk will give the correct $n$-point functions of the
free theory on the boundary.

\section{Algebraic structure of the higher spin symmetries.}
\subsection{Oscillator representation of $su(2,2)$.}
The higher spin symmetry algebra is an infinite dimensional extension
of the conformal algebra $su(2,2)$ \cite{hs}. 
The simplest description of this
algebra is in terms of oscillators \cite{Oscillators}. 
The construction is based on the embedding of the unitary algebras
into symplectic algebras. For example
$su(2,2)$ is a subalgebra of $Sp(4,{\bf R})$. Let us review the
construction of the embedding.

Let $V$ be the 8-dimensional ${\bf R}$-linear space with
the symplectic structure, that is with the nondegenerate two-form
$\omega$.
The quadratic hamiltonians generate $sp(4,{\bf R})$.
Suppose that $V$ is also equipped with the complex structure
$I$, so that $\omega$ is of the type $(1,1)$. The action of the
complex structure on $V$ is generated by the quadratic hamiltonian
which we will call $H_c$. Let us consider those quadratic hamiltonians
which commute with $H_c$. These hamiltonians form an algebra, which is
in fact $u(M,4-M)$.
Here $M$ depends on how
we choose the complex structure. It is the number of positive eigenvalues
of the symmetric form $g(v_1,v_2)=\omega(v_1,I.v_2)$.
Let us choose such a complex structure that $M=2$. This choice is
characterized by the existence of the holomorphic Lagrangean
subspaces in $V$. In other words, in the complexification of $V$
we can choose coordinates $Q^I$ and $P_I$ so that
\begin{equation}
[H_c,P^I ] = iP^I,\; {[} H_c,Q^I {]} = iQ^I
\end{equation}
and
\begin{equation}
\begin{array}{l}
[P^I,Q^J] = [P^I,P^J] = [Q^I,Q^J] =0,\\[5pt]
[P^I,\overline{P}^J]=[Q^I,\overline{Q}^J]=0,\\[5pt]
[P^I,\overline{Q}^J]=\delta^{IJ}
\end{array}
\end{equation}
There is an invariant bilinear form on $u(M,4-M)$ which comes from the
invariant bilinear form on $sp(4,{\bf R})$:
\begin{equation}
||H||^2=(H_{Weyl})^2-(H^2)_{Weyl}
\end{equation}
where the subscript means the Weyl ordering of the oscillators (the sum
over all the permutations.)
The algebra $u(2,2)$ has a center which is generated by $H_c$.
The orthogonal complement to this center with respect to the invariant
bilinear form is $su(2,2)$.

\subsection{Higher Spin Algebra.}
The algebra $hs(2,2)$ is the infinite dimensional algebra generated by
all the hamiltonians which commute with $H_I$ (not only quadratic).
This is an associative algebra, but we will need it as a Lie algebra.
The generic element of $hs(2,2)$ has the following form:
\begin{equation}\label{hsTwoTwo}
\underline{hs(2,2)}:
f(a,\bar{a})=\alpha+\alpha_{i\bar{j}}a^i\bar{a}^{\bar{j}}
+\alpha_{i_1i_2\bar{j}_1\bar{j}_2}
a^{i_1}a^{i_2}\bar{a}^{\bar{j}_1}\bar{a}^{\bar{j}_2}
+\ldots
\end{equation}
where $a$ and $\overline{a}$ are elements of $V_{\bf C}^{(1,0)}$
and $V_{\bf C}^{(0,1)}$, respectively. We use complex notations
for oscillators, but
we should stress that this is a real algebra. All the terms
in (\ref{hsTwoTwo}) are manifestly real if we use real oscillators.
As a Lie algebra this algebra has an outer ${\bf Z}_2$-automorphism:
\begin{equation}
{\bf i}.f(a,\bar{a})=
-f(-a,\bar{a})
\end{equation}
In this formula we assume the Weyl ordering of the oscillators. 
This is an automorphism of the Lie algebra
because the symplectic structure is of the type $(1,1)$.
The subalgebra of $hs(2,2)$ invariant under this automorphism will
be denoted $hs_{\bf R}(2,2)$. This subalgebra contains $su(2,2)$.
Its generic element has the following form:
\begin{equation}
\underline{hs_{\bf R}(2,2)}:
\;\;\;\;
f=\alpha_{i\bar{j}}a^i\bar{a}^{\bar{j}}
+\alpha_{i_1i_2i_3\bar{j}_1\bar{j}_2\bar{j}_3}
a^{i_1}a^{i_2}a^{i_3}\bar{a}^{\bar{j}_1}\bar{a}^{\bar{j}_2}
\bar{a}^{\bar{j}_3}+\ldots
\end{equation}
It is not an associative algebra.

Let us consider the complexification $hs(2,2,{\bf C})=
{\bf C}\otimes hs(2,2)$. Consider in $hs(2,2,{\bf C})$ the following
reality condition:
\begin{equation}
{\bf i}.f=\bar{f}
\end{equation}
The real subalgebra will be denoted $hs_{\bf C}(2,2)$. It consists of the
following elements:
\begin{equation}\label{hsC}
\underline{hs_{\bf C}(2,2)}:\;\;\;\;
f=i\alpha+\alpha_{i\bar{j}}a^i\bar{a}^{\bar{j}}
+i\alpha_{i_1i_2\bar{j}_1\bar{j}_2}
a^{i_1}a^{i_2}\bar{a}^{\bar{j}_1}\bar{a}^{\bar{j}_2}
+\ldots
\end{equation}
where all the coefficients $\alpha$ are hermitean matrices.
Again, this $hs_{\bf C}(2,2)$ is not an associative algebra, only a Lie
algebra.

We will see that $hs_{\bf R}(2,2)$ is the algebra of symmetries of
the free real scalar fields, and $hs_{\bf C}(2,2)$ is the algebra
of symmetries of the free complex field which respect the complex
structure (defined over $\bf C$.)

\section{Free fields as doubletons.}
\subsection{Definition.}
Consider the operators in the free scalar field theory (on the boundary)
which are linear in the free field. For example, 
$\partial_i \partial_j\; \phi(x)$ is such an operator.
The space of all such operators is the representation of the
higher spin symmetry algebra. It is called "doubleton representation". 
It was introduced (as a representation of the 
conformal algebra) in \cite{Singleton,Doubleton}. 
In this section we will review the construction of this representation.

{\em Doubleton representation of $\hsC$.}
Let us consider the representation of the oscillators on the space
of complex functions $\psi(q,\bar{q})$
where $P$ and $Q$ act in the following way:
\begin{equation}\label{OscRep}
Q^I.\psi(q,\bar{q})=q^I\psi(q,\bar{q}),\;\;\;
\overline{P}^I.\psi(q,\bar{q})=
{\partial\over\partial q^I}\psi(q,\bar{q})
\end{equation}
This gives a representation  of the associative algebra
$hs(2,2)$. As a representation of $hs(2,2)$ it contains an
invariant subspace consisting of those functions which are invariant
under $(q,\bar{q})\to (e^{i\theta}q, e^{-i\theta}\bar{q})$.
This invariant subspace is the
doubleton representation of $\hsC$, we will call it $\cal F$.
We will consider this $\cal F$ as a representation of the
Lie algebra $hs_{\bf C}(2,2)$.

\noindent
{\em Doubleton representation of $hs_{\bf R}(2,2)$.}
Let us now consider this representation as the representation of
the subalgebra $hs_{\bf R}(2,2)\subset hs_{\bf C}(2,2)$.
As a representation of $hs_{\bf R}(2,2)$ it has an invariant
subspace consisting of the real functions $\psi(q,\bar{q})$.
This subspace is an irreducible representation of $hs_{\bf R}(2,2)$.

\subsection{Doubletons as free fields. Scalar product.}
The doubleton representation $\cal F$ of $hs_{\bf C}(2,2)$ has an
invariant hermitean scalar product:
\begin{equation}\label{SPNaive}
(\psi_1,\psi_2)=\int d^4 q \;\overline{\psi_1(-q,\overline{q})}
\psi_2(q,\overline{q})
\end{equation}
This expression is somewhat formal, because we did not specify which
functions we need, and there is no guarantee that the integral will
converge. Let us consider a special class of functions, of the following
form:
\begin{equation}\label{Laplace}
\psi(q,\bar{q})=\int d^4 x \exp\left(
-x_{I\bar{J}}q^Iq^{\bar{J}}\right) \phi(x)
\end{equation}
where $\phi(x)$ is some function of $x$, say with compact support.
This formula tells us that the doubleton representation is
the space of functions (rapidly decreasing, or with compact support)
$\phi(x)$ modulo those functions which are laplacians of the function
with the compact support:
\begin{equation}
\phi(x)\sim \phi(x)+\Delta \chi(x)
\end{equation}
Suppose that we have two functions $\phi_1(x)$ and $\phi_2(x)$
such that the support of $\phi_2(x)$ is in the future of the
support of $\phi_1(x)$.
Then the integral
in (\ref{SPNaive}) converges and is equal to:
\begin{equation}\label{SPDefinition}
(\phi_1,\phi_2)=\int d^4x_1 d^4x_2 {\phi_1^*(x_1)\phi_2(x_2)\over
||x_1-x_2||^2}
\end{equation}
This is the invariant hermitean scalar product of $\phi_1$ and $\phi_2$.
It is equal to the correlation function
\begin{equation}
\langle \int d^4 x \phi_1^*(x)\varphi(x)
\int d^4y \phi_2(x)\varphi^*(x) \rangle
\end{equation}
where $\varphi(x)$ is the free scalar field with the lagrangian
$\int d^4x \partial_{\mu}\varphi^*(x) \partial^{\mu}\varphi(x)$.

The doubleton representation of $hs_{\bf R}(2,2)$ also
 has an orthogonal scalar product defined by the same formula
(\ref{SPDefinition}) but with the real functions $\phi_1$ and
$\phi_2$.

The action of $\hsC$ on $\phi(x)$ can be read from
the action of $\hsC$ on $\psi(q,\bar{q})$.
One can see that $\hsC$ acts on $\phi(x)$ by the differential
operators with the coefficients depending on $x$. In particular,
all the differential operators with constant coefficients
are elements of $\hsC$. (They are the
polynomials of $Q_IQ_{\overline{J}}$.)

The doubleton representation has a simple geometrical meaning.
The points of the four dimensional Minkowski space may be thought
of as holomorphic Lagrangean planes in the space of oscillators.
The insertion of the operator $\phi$ at the point $x$ corresponds
to the vector of the Fock space annihilated by all the oscillators
belonging to the Lagrangean plane which corresponds to the point $x$.
For example to the point $x=0$ corresponds the wave function
$\psi(q,\bar{q})=1$ which is annihilated by all the momenta $P^I$.
The correlation functions $<\phi^*(x)\; \phi(y)>$ corresponds to the
scalar product of the vectors corresponding to the points $x$ and
$y$.

\noindent
{\em $\hsC$ is the algebra of conformal Killing tensors.}
Suppose that we act on $\phi$ by the differential operator, polynomial
in $x$ and $\partial_x$. Integrating by parts in (\ref{Laplace})
we see that the action on $\exp(x_{I\bar{J}}q^Iq^{\bar{J}})$ is by
the multiplication by the polynomial in ${\cal P}(x, q\bar{q})$.
Our differential operator is a symmetry, therefore
\begin{equation}
{\partial\over\partial x^{\mu}}
{\partial\over\partial x_{\mu}}\;
\left({\cal P}(x,q\bar{q})\exp(x_{I\bar{J}}q^Iq^{\bar{J}})\right)=0
\end{equation}
This implies that
\begin{equation}\label{WhatWeNeed}
{\cal P}(x,q\bar{q})\exp(x_{I\bar{J}}q^Iq^{\bar{J}})=
{\cal P}'(q,\bar{q},\partial_q,\partial_{\bar{q}})
\exp(x_{I\bar{J}}q^Iq^{\bar{J}})
\end{equation}
where ${\cal P}'$ is some polynomial. We give the prove of this
statement in the Appendix.

A higher derivative symmetry of the Laplace equation
$\Delta \phi=0$ is related to some conformal Killing tensor.
Indeed, for the differential operator to be a symmetry of the
Laplace equation $\Delta\phi=0$ it is necessary that the leading symbol
of the operator (which is the coefficient of the highest power of
the derivative) is a conformal Killing tensor.
Therefore we have proven that $\hsC$ is the algebra of conformal
Killing tensors.

{\em Conformal Killing tensors are the linear combinations
of the products of the conformal Killing vectors.}
Indeed, we have established the correspondence between the
conformal Killing vectors and $I$-invaiant polynomials of oscillators.
But any $I$-invariant polynomial of oscillators can be rewritten
as the linear combination of the products of the bilinears.
This implies that any symmetry is a linear combination of the products
of the conformal trnasformations.

It is unusual that the product of the symmetry generators is again
a generator of the symmetry. This happens because we are dealing with
the free fields. The differential operator $L$ being a symmetry means
that:
\begin{equation}
 \int \partial_{\mu}\phi^* \partial^{\mu}(L.\phi)=
-\int \partial_{\mu}(L.\phi)^*\partial^{\mu}\phi
\end{equation}
Then, for $k$ such symmetries the operator
$i^{k+1}(L_1\cdots L_k+L_k\cdots L_1)$
is again a symmetry:
\begin{equation}
\begin{array}{l}
 \int \partial_{\mu}\phi^* \partial^{\mu}
(i(L_1\cdots L_k+L_k\cdots L_1).\phi)=\\[5pt]=
-\int \partial_{\mu}(i(L_1\cdots L_k+L_k\cdots L_1).\phi)^*
\partial^{\mu}\phi
\end{array}
\end{equation}
Therefore there is a multiplication on the space of
conformal Killing tensors. This multiplication is, of course,
just a tensor product of two conformal Killing tensors.
Notice that in the bulk, there is no obvious multiplicative
structure on the traceless Killing tensors (the product of two
traceless Killing tensors is again a Killing tensor, but not
traceless.)

{\em Nonlocal symmetries.}
All the higher spin symmetries can be obtained from some nonlocal
symmetries. Let us fix a pair of points $(z,w)$ in ${\bf R}^4$.
To every such pair corresponds a nonlocal symmetry of the free action
which acts on the free fields as follows:
\begin{equation}
\begin{array}{l}
\delta_{(z,w)}\phi(x)={1\over ||x-w||^2}\phi(z)\\
\delta_{(z,w)}\phi^*(x)=-{1\over ||x-w||^2}\phi^*(w)
\end{array}
\end{equation}
One can see that this is a symmetry of the free action.
This transformation as we defined it does not respect $\phi^*$
being complex conjugate to $\phi$. But this can be fixed
by considering the linear combinations $\delta_{(z,w)}-\delta_{(w,z)}$
and $i(\delta_{(z,w)}+\delta_{(w,z)})$ instead of $\delta_{(z,w)}$.
These nonlocal symmetries may be thought of as infinite
linear combinations of the higher spin symmetries.

\subsection{Constraints imposed on the correlation functions by the
higher spin symmetries.}
Let us consider the tensor product ${\cal F}^{\otimes n}$.
All the operators in ${\cal F}^{\otimes n}$
commuting with $\hsC$ are linear combinations of the permutations.

To explain why, we need the following fact: all the linear
operators in ${\cal F}$ are ${\bf C}$-linear combinations
of the generators of $\hsC$. We are including infinite sums and
integrals over
continuous parameters into the notion of ``linear combinations.
We will not address the questions of convergence.
The space ${\cal F}$ is the space of all the functions $\psi(q^i)$,
$i\in\{1,\ldots,4\}$ invariant under $I$. We will restrict ourselves
to the space of polynomials $\psi(q^i)$.
Let us introduce the ``multiindex'' notation for the monomials:
\begin{equation}
\vec{q}^{\vec{m}}=(q^1)^{m_1}(q^2)^{m_2}(\bar{q}^1)^{m'_1}
(\bar{q}^2)^{m'_2},\;\;\; m_1+m_2=m'_1+m'_2
\end{equation}
We will denote
\begin{equation}
X(\alpha_1,\alpha_2,\alpha'_1,\alpha'_2)=
\exp \left[ i\alpha_1 q^1{\partial\over\partial q^1}+
 i\alpha_2 q^2{\partial\over\partial q^2}+
i\alpha'_1 \overline{q}^1{\partial\over\partial \overline{q}^1}+
i\alpha'_2 \overline{q}^2{\partial\over\partial \overline{q}^2}\right]
\end{equation}
If we decompose this $X$ in powers of $\alpha$ we see that it is
an infinite series of the generators of the higher spin algebra
acting as in (\ref{OscRep}).
Then, for the two given multiindices $\vec{a}$ and $\vec{b}$
the operator
\begin{equation}
X^{\vec{a}}_{\vec{b}}:={1\over b_1! b_2!b'_1!b'_2!}
\int d^4\alpha e^{-i\sum\alpha_jb_j}\;
q^{\vec{a}}\left({\partial\over\partial q}\right)^{\vec{b}}
X(\alpha_1,\alpha_2,\alpha'_1,\alpha'_2)
\end{equation}
is manifestly a linear combination (in fact, a continuous integral)
with the complex coefficients of the generators of $\hsC$. On the
other hand, its action on monomials is:
\begin{equation}
X^{\vec{a}}_{\vec{b}}.q^{\vec{c}}=
\delta^{\vec{c}}_{\vec{b}}q^{\vec{a}}
\end{equation}
An arbitrary linear operator acting on the space of polynomials
can be represented as a linear combination of such operators.

This means that in the tensor product
${\cal F}\otimes\cdots\otimes{\cal F}$ any
linear operator $X:\;{\cal F}\to{\cal F}$ acting as
\begin{equation}\label{Coproduct}
\begin{array}{l}
X.(v_1\otimes v_2\otimes\cdots\otimes v_n)=\\[5pt]=
Xv_1\otimes v_2\otimes\cdots\otimes v_n+
v_1\otimes X v_2\otimes\cdots\otimes v_n+\ldots+
v_1\otimes v_2\otimes\cdots\otimes X v_n
\end{array}
\end{equation}
is a linear combination of the generators of the higher spin algebra.
Therefore the operator commuting with the higher spin symmetries
should commute with any linear operator acting as (\ref{Coproduct}).
But if the operator in the tensor product space commutes with
any linear operator acting as (\ref{Coproduct}), then it is a
permutation operator.

Operators in ${\cal F}^{\otimes n}$ commuting with the higher spin
algebra are the same as  invariants of the higher spin symmetry
in the space ${\cal F}^{\otimes n}\otimes {\cal F}^{*\otimes n}$.
We see that all such invariants are parameterized by the permutations
$\sigma\in S_n$:
\begin{equation}
\langle v_1\otimes\cdots\otimes v_n,\;
        w_1\otimes\cdots\otimes w_n \rangle_{\sigma}=
(v_1\cdot w_{\sigma_1})\cdots(v_n\cdot w_{\sigma_n})
\end{equation}
This statement may be reformulated as follows.
Suppose that we are given a four
dimensional theory which enjoys the higher spin symmetry $hs(2,2)_{\bf C}$.
Suppose that the local operators in this theory can be described as
tensors $T^{i_1\ldots i_p;\; j_1\ldots j_q}(x)$ traceless and symmetric
in both $i$ indices and $j$ indices
and the transformation
of this tensors under the higher spin symmetry is the same as the
transformation of the operator
$$T^{i_1\ldots i_p;\;j_1\ldots j_q}\;
\partial_{i_1}\cdots \partial_{i_p}\phi(x)\;
\partial_{j_1}\cdots\partial_{j_q}\phi^*(x)$$
in the theory of the free complex scalar $\phi(x)$. (For example,
the translation acts as follows. The action of the translation $P_k$
on the tensor $T^{i_1\ldots i_p;\; j_1\ldots j_q}$ is the
sum of the two tensors of the rank $(p+1,q)$ and $(p,q+1)$:
$\delta_k^{i_1}T^{i_2\ldots i_{p+1};\;j_1\ldots j_q}+
 \delta_k^{j_1}T^{i_1\ldots i_p;\; j_2\ldots j_{q+1}}$.)
Then, the most general form of the correlation functions of these
operators compatible with the higher spin symmetries
is:
\begin{equation}\label{Ambiguity}
\begin{array}{l}
\langle T_{i_{1,1}\ldots i_{1,p};\;j_{1,1}\ldots j_{1,q}}(x_1)\cdots
        T_{i_{n,1}\ldots i_{n,p};\;j_{n,1}\ldots j_{n,q}}(x_n)
\rangle
=\\[5pt]=
T_{i_{1,1}\ldots i_{1,p_1};\;j_{1,1}\ldots j_{1,q_1}}(x_1)\cdots
        T_{i_{n,1}\ldots i_{n,p_n};\;j_{n,1}\ldots j_{n,q_n}}(x_n)
\times\\[5pt]\times
\sum\limits_{\sigma\in {\cal S}_n}A_{\sigma}
\prod\limits_{k=1}^n
<\partial_{j_{\sigma(k),1}}\cdots
\partial_{j_{\sigma(k),q_{\sigma(k)}}}
\phi^*(x_{\sigma(k)})\;
\partial_{i_{k,1}}\cdots \partial_{i_{k,p_k}}\phi(x_k)>_{free}
\end{array}
\end{equation}
where the correlators on the right hand side are taken in the
free scalar field theory.
All the ambiguity not fixed by the higher spin symmetries is in
$A_{\sigma}$.

We have seen in Section 3 that the free higher spin fields in the bulk
correspond to the primary fields on the boundary which are the
traceless tensors with zero divergence. Adding the descendants, we
get tensors $T^{i_1\ldots i_p;\; j_1\ldots j_q}$ which are
traceless and symmetric in both $i$ and $j$.
If the two conditions formulated in Section 4.2 are satisfied
then the action of the higher spin symmetries on
$T^{i_1\ldots i_p;\; j_1\ldots j_q}$ agrees with the action on the
free fields. Then the general $n$-point function (given by the
complicated multiple integrals over the AdS space) will necessarily
be of the general form (\ref{Ambiguity}). In particular, it is
a rational function of $x_1,\ldots,x_n$. All the ambiguity for the $n$
point function is in $n!$ coefficients $A_{\sigma}$, and it is
presumably fixed by the positions of the singularities.

\vspace{25pt}
\hspace{-21.5pt}
{\Large \bf Acknowledgements}\vspace{12pt}\\
I want to thank E.~Witten for drawing my attention to the problem
of higher spin interactions and for discussions.
I enjoyed discussions with M.~Berkooz, D.~Gross, A.~Jevicki, S.~Minwalla,
A.M.~Polyakov, E.~Sezgin, K.~Skenderis and N.~Toumbas.
I want to thank the Theory Groups of the University of Amsterdam,
Harvard University and the Institute for Advanced Study
for their hospitality.
This work was supported in part by the NSF Grant No. PHY99-07949,
and in part by RFBR Grant No. 00-02-16477 and by the
Russian grant for the support of the scientific schools No. 00-15-96557.

\appendix
\section{Technical lemma.}
We want to prove that for any symmetric traceless tensor
$f^{\mu_3\ldots \mu_s}$ there is a symmetric traceless
$\Lambda^{\mu_2\ldots \mu_s}$ such that
\begin{equation}\label{ExistLambda}
f^{\mu_3\ldots\mu_s}=\nabla_{\rho}\Lambda^{\rho\mu_3\ldots\mu_s}
\end{equation}
The AdS space is conformally flat, which means that the metric
can be written as $g_{\mu\nu}=\phi^2 \delta_{\mu\nu}$. In flat
coordinates
\begin{equation}
\nabla_{\rho} \Lambda^{\rho\mu_3\ldots\mu_s}=
\phi^{d+2(s-2)\over 2}\partial_{\rho}
\left(\phi^{-{d+2(s-2)\over 2}}\Lambda^{\rho\mu_3\ldots\mu_s}
\right)
\end{equation}
Therefore it is enough to prove (\ref{ExistLambda}) in flat space.
In flat space we use the Fourier transform:
$$
f^{\mu_3\ldots\mu_s}(x)=\int d^d k
e^{ikx}\hat{f}^{\mu_3\ldots\mu_s}(k)
$$
We have to prove that for any $\hat{f}^{\mu_3\ldots\mu_s}$
there is $\hat{\Lambda}^{\mu_2\ldots\mu_s}$ symmetric and traceless,
such that
\begin{equation}\label{Fourier}
k_{\rho}\hat{\Lambda}^{\rho\mu_3\ldots\mu_s}=
\hat{f}^{\mu_3\ldots\mu_s}
\end{equation}
There is an obvious correspondence between the rank $r$ symmetric
tensors
in $d$ dimensions and the degree $r$ monomials in $d$ variables
$u_1,\ldots,u_d$. Traceless symmetric tensors correspond to
harmonic polynomials.
If $P_f$ is the monomial corresponding to $f$ and $P_{\Lambda}$
is the monomial corresponding to $\Lambda$ then (\ref{Fourier})
becomes:
\begin{equation}\label{DivForPolynomials}
P_f(u)=k_{\rho}{\partial\over\partial u_{\rho}}P_{\Lambda}(u)
\end{equation}
By the change of variables we can have $k_1=k$ and $k_{\mu}=0$
for $\mu>1$. Then, a solution for (\ref{DivForPolynomials})
is
\begin{equation}
P_{\Lambda}(u_1,\ldots,u_d)={1\over k}\left[\int_0^{u_1} dv
P_f(v,u_2,\ldots,u_d)-Q(u_2,\ldots,u_d)\right]
\end{equation}
where $Q(u_2,\ldots,u_d)$ is a solution to the equation:
\begin{equation}
(\partial^2_{u_2}+\ldots+\partial^2_{u_d})Q(u_2,\ldots,u_d)=
\partial_{u_1}P_f(u_1,\ldots,u_d)+
\int_0^{u_1} dv
(\partial^2_{u_2}+\ldots+\partial^2_{u_d})P_f(v,u_2,\ldots,u_d)
\end{equation}
Notice that the right hand side of this equation does not depend
on $u_1$ (because $P_f(u)$ is a harmonic polynomial.)
\section{Another technical lemma.}
Suppose that $P(q,\bar{q},x)$ is a polynomial (of finite degree)
in all its variables, and the function $P(q,\bar{q},x)e^{q\bar{q}x}$
is harmonic:
\begin{equation}\label{LaplaceOnPExp}
\Delta \left( P(q,\bar{q},x)e^{q\bar{q}x} \right)=0
\end{equation}
Then, there is a polynomial
$\tilde{P}(q,\bar{q},\partial_q,\partial_{\bar{q}})$ such that
\begin{equation}
P(q,\bar{q},x)e^{q\bar{q}x}=
 \tilde{P}(q,\bar{q},\partial_q,
\partial_{\bar{q}})e^{q\bar{q}x}
\end{equation}
Let us prove it. We will need the equation for $P$ which is equivalent
to the Laplace equation for $P\exp(q\bar{q}x)$ (\ref{LaplaceOnPExp}):
\begin{equation}\label{LaplaceOnP}
\left( \Delta + \epsilon_{II'}\epsilon_{\bar{J}\bar{J}'}
q^{I'}\bar{q}^{\bar{J}'} {\partial\over\partial x_{I\bar{J}}}
\right)P(q,\bar{q},x)=0
\end{equation}
Let us take the terms in $P$ which are of the highest degree in
$q\bar{q}$. We will call these terms $P_1$:
\begin{equation}
P=P_1+\mbox{lower powers of }\;q\bar{q}
\end{equation}
It follows from (\ref{LaplaceOnP}) that
\begin{equation}
\epsilon_{II'}\epsilon_{\bar{J}\bar{J}'}
q^{I'}\bar{q}^{\bar{J}'} {\partial\over\partial x_{I\bar{J}}}
P_1(q,\bar{q},x)=0
\end{equation}
This implies that
\begin{equation}\label{PThroughf}
P_1(q,\bar{q},x)=
f(q,\bar{q},q^I x_{I\bar{J}}, \bar{q}^{\bar{J}}x_{I\bar{J}})
\end{equation}
with some polynomial $f$.
Let us define the polynomial $\tilde{P}$ from the formula:
\begin{equation}
P(q,\bar{q},x)\exp(q\bar{q}x)=
f(q,\tilde{q},\partial_{\bar{q}},\partial_q).\exp(q\bar{q}x)
+\tilde{P}(q,\bar{q},x)\exp(q\bar{q}x)
\end{equation}
Notice that $\tilde{P}$ has lower degree in $q\tilde{q}$ than
$P$. Also, $\tilde{P}\exp(q\bar{q}x)$ satisfies the Laplace
equation just as $P\exp(q\bar{q}x)$.
Now we can play the same game with $\tilde{P}$:
take the part of
$\tilde{P}$ which has the highest degree in $q\bar{q}$,
represent it in the form (\ref{PThroughf}) with some $\tilde{f}$,
and then define $\tilde{\tilde{P}}$, which has smaller degree in
$q\bar{q}$ than $\tilde{P}$.
After repeating this procedure finitely many times, we arrive
at the representation
\begin{equation}
P(q,\bar{q},x)\exp(q\bar{q}x)=
(f(q,\bar{q},\partial_{\bar{q}},\partial_q)+
\tilde{f}(q,\bar{q},\partial_{\bar{q}},\partial_q)+\ldots)
\exp(q\bar{q}x)
\end{equation}
which is what we need.

\end{document}